\documentclass{article}
\usepackage{amsmath}
\usepackage{graphics}
\usepackage{color}

\input{tcilatex}

\begin{document}

\title{Heisenberg's Introduction of the `Collapse of the Wavepacket' into
Quantum Mechanics}
\author{Raymond Y. Chiao \\
Department of Physics\\
University of California\\
Berkeley, CA 94720-7300\\
U. S. A.\\
E-mail: chiao@physics.berkeley.edu \and Paul G. Kwiat \\
Department of Physics\\
University of Illinois\\
Urbana, IL 61801\\
U. S. A.\\
E-mail: kwiat@uiuc.edu}
\date{January 23, 2002}
\maketitle

\begin{abstract}
Heisenberg in 1929 introduced the ``collapse of the wavepacket'' into
quantum theory. \ We review here an experiment at Berkeley which
demonstrated several aspects of this idea. \ In this experiment, a pair of
daughter photons was produced in an entangled state, in which the sum of
their two energies was equal to the sharp energy of their parent photon, in
the nonlinear optical process of spontaneous parametric down-conversion. \
The wavepacket of one daughter photon collapsed upon a
measurement-at-a-distance of the other daughter's energy, in such a way that
the total energy of the two-photon system was conserved. \ Heisenberg's
energy-time uncertainty principle was also demonstrated to hold in this
experiment.
\end{abstract}

\section{Introduction}

In this Symposium in honor of Heisenberg's Centennial, it is appropriate to
begin by recalling the fact that in the spring of 1929, during his lectures
at the University of Chicago, Heisenberg introduced the important concept of
the ``collapse of the wavepacket'' into quantum theory \cite{Heisenberg1929}%
. \ This idea, which he referred to as the ``reduction of the wavepacket,''
was closely related to the idea of the ``collapse of the wavefunction,''
which was introduced into the standard Copenhagen interpretation of quantum
mechanics in connection with the probabilistic interpretation of the
wavefunction due to Born \cite{Born}. \ In the context of a remark
concerning the spreading of the wavepacket of an electron, Heisenberg stated
the following \cite{Heisenberg1929}:

\begin{quote}
In relation to these considerations, one other idealized experiment (due to
Einstein) may be considered. \ We imagine a photon which is represented by a
wave packet built up out of Maxwell waves \cite{HeisenbergFootnote}. \ It
will thus have a certain spatial extension and also a certain range of
frequency. \ By reflection at a semi-transparent mirror, it is possible to
decompose it into two parts, a reflected and a transmitted packet. \ There
is then a definite probability for finding the photon either in one part or
in the other part of the divided wave packet. \ After a sufficient time the
two parts will be separated by any distance desired; now if an experiment
yields the result that the photon is, say, in the reflected part of the
packet, then the probability of finding the photon in the other part of the
the packet immediately becomes zero. \ The experiment at the position of the
reflected packet thus exerts a kind of action (reduction of the wave packet)
at the distant point occupied by the transmitted packet, and one sees that
this action is propagated with a velocity greater than that of light. \
However, it is\ also obvious that this kind of action can never be utilized
for the transmission of signals so that it is not in conflict with the
postulates of the theory of relativity.\ 
\end{quote}

At the heart of Heisenberg's (and, earlier, Einstein's) conception of the
``collapse of the wavepacket,'' was the $indivisibility$ of the individual
quantum, here of the light quantum -- the photon -- at the beam splitter %
\cite{Newton}. \ Ultimately, it was the indivisibility of the photon that
enforced the collapse of the wavepacket, whenever a detection of the
particle occurred at one of the two exit ports of the beam splitter (for
example, whenever a ``click'' occurred at a counter placed at the reflection
port); because the photon was $indivisible$, it had $either$ to be
reflected, $or$ to be transmitted at the beam splitter, but not both. \
Remarkably, a \textit{non}-detection by a 100\%-efficient detector will also
collapse the state, so that the photon is definitely in the other arm \cite%
{Dicke}. \ 

Heisenberg believed that the reduction or collapse of the wavepacket was
indeed a physical $action$, and not just a convenient $fiction$ which was
useful in the interpretation of quantum phenomena but which had no physical
reality\ attached to it. \ It is an irony of history that although
Heisenberg attributed this idea to Einstein, it was later totally rejected
by Einstein along with all of its consequences, as unexplained, ``spooky
actions-at-a-distance (\textit{spukhafte Fernwirkungen}).'' The
``collapse''\ idea, and its later developments, culminated in Einstein's
ultimate rejection of quantum theory as being an incomplete theory of the
physical world.

In contrast, von Neumann embraced the idea, and further sharpened it by
introducing the notion of ``projection of rays in Hilbert space''\ upon
measurement, in the last two chapters entitled `Measurement and
Reversibility' and `The Measuring Process' of his book \textit{Mathematical
Foundations of Quantum Mechanics }\cite{Von Neumann}. \ The physics of the
``collapse postulate'' introduced by Heisenberg was thereby tied to the
mathematics of the ``von Neumann projection postulate.'' \ In this way, the
Copenhagen interpretation of quantum mechanics, or at least one version of
it, was completed.

\begin{figure}[tbp]
\centerline{\includegraphics{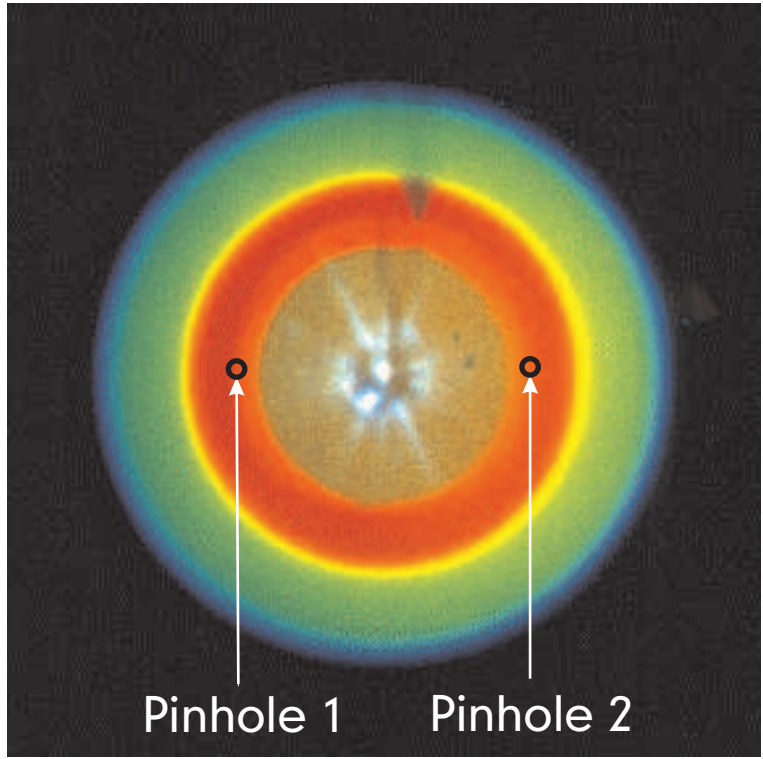}}
\caption{Color photograph (an end-on view) of the spontaneous parametric
down-conversion from an ultraviolet\ ($\protect\lambda =$ 351 nm) laser beam
which has traversed a KDP crystal. \ (Some of the UV laser light, which has
leaked through an UV rejection filter, caused the irregular splotches near
the center). By means of Pinhole 1 and Pinhole 2, correlated pairs of
photons emitted near the same red wavelength ($\protect\lambda =$ 702 nm)
were selected out for coincidence detection.\ \ This picture illustrates yet
another striking aspect of the ``collapse'' idea, which is different from
the one discussed in the text. \ Although the angular distribution of the
probability amplitude for the emission of the photon pair starts off as an
azimuthally isotropic ring around the center, once a ``click'' is registered
by a detector placed behind one of the pinholes, there is the sudden onset
of a type of ``spontaneous symmetry breaking,'' in which the probability for
detection of the other photon suddenly vanishes everywhere around the ring,
except at the other pinhole, where the probability of finding it suddenly
becomes unity. \ (Adapted from A. Migdall and A. V. Sergienko's original
photo).}
\label{fig1}
\end{figure}

Here we shall review an experiment performed at Berkeley on another aspect
of the ``collapse of the wavepacket,'' as viewed from the standard
Copenhagen viewpoint. \ This experiment involved a study of the\ correlated
behaviors of two photon wavepackets in an entangled state\ of energy. \ The
two entangled photon wavepackets were produced in the process of spontaneous
parametric down-conversion, in which a parent photon from an ultraviolet
(UV) laser beam was split inside a nonlinear crystal into two daughter
photons, conserving energy and momentum during this process. \ One can view
this quantum nonlinear optical process as being entirely analogous to a
radioactive decay process in nuclear physics, in which a parent nucleus
decays into two daughter nuclei. \ We shall see that a measurement of the
energy of one daughter photon has an instantaneous collapse-like
action-at-distance upon the behavior of the other daughter photon.\

\section{The entangled photon-pair light source: Spontaneous parametric
down-conversion}

We produced pairs of energy-entangled photon wave packets by means of
spontaneous parametric down-conversion, also known as ``parametric
fluorescence,'' in an optical crystal with a nonvanishing $\chi ^{(2)}$
nonlinear optical susceptibility \cite{Burnham}. \ In our experiment, we
employed a crystal of potassium dihydrogen phosphate (KDP) \cite{Kwiat1991}.
\ The lack of inversion symmetry in crystals such as KDP breaks the usual
parity-conservation selection rule, so that it is not forbidden for one
photon to decay into two photons inside the crystal. \ 

There are many ways in which a single, monoenergetic, parent photon
(conventionally called the ``pump'' photon, originating in our experiment
from an argon ion UV laser operating at $\lambda =$ 351 nm) can decay into
two daughter photons (conventionally called the ``signal'' and the ``idler''
photons), while distributing its energy between\ the two in a continuous
fashion. \ There is\ therefore no reason why the daughter photons would
necessarily be monoenergetic. \ In fact, as a result of the normal
dispersion in the linear refractive index of the crystal, it turns out that
the conservation of energy and momentum in the two-photon decay process
results in the production of a rainbow of conical emissions of photon pairs
with a wide spectrum of colors, which is shown in Figure 1. \ Two photons on
diametrically opposite sides of the rainbow are emitted in a pairwise
fashion, conserving energy and momentum in the emission process. \ 

By means of two pinholes, we selected out\ of the rainbow for further study
two tightly correlated, entangled photons, which were emitted around the
same red wavelength (i.e., $\lambda =$ 702 nm, at twice the wavelength of
the pump photon). \ For millimeter-scale pinholes, which span a few percent
of the full visible spectrum, the resulting photon wavepackets typically
have subpicosecond widths \cite{CWpump}. \ Thus, two photon counters placed
behind these two pinholes would register tightly correlated, coincident
``clicks.''

The\ KDP crystal which we used was 10 cm long, cut such that its c-axis was
50.3$^{\mathrm{o}}$ to the normal of its input face; the UV laser beam was
normally incident on the KDP crystal face, with a vertical linear
polarization. \ The correlated signal and idler photon beams were both
horizontally polarized. \ The two pinholes were placed at $+1.5^{\mathrm{o}}$
and at $-1.5^{\mathrm{o}}$ with respect to the UV laser beam, so that
degenerate pairs of photons centered at a wavelength of 702 nm were selected
for study. \ Thus, in this parametric fluorescence process, a single parent
photon with a sharp spectrum from the UV laser was spontaneously converted
inside the crystal into a pair of daughter photons with broad, conjugate
spectra centered at half the parent's UV energy.

The photon pair-production\ process due to parametric down-conversion
produces the following entangled state \cite{OmitVacuum}:%
\begin{equation}
\left| \psi \right\rangle =\int dE_{1}A(E_{1})\left| 1\right\rangle
_{E_{1}}\left| 1\right\rangle _{E_{2}=E_{0}-E_{1}}\text{ ,}
\end{equation}%
where $E_{0}$ is the energy of the parent UV photon,\ $E_{1}$ is the energy
of the first member of the photon pair (the ``idler'' daughter photon), $%
E_{2}=E_{0}-E_{1}$ is the energy of the second member of the photon pair
(the ``signal'' daughter photon), and $A(E_{1})$ is the probability
amplitude for the emission of the pair. \ The first (second) photon is in
the one-photon Fock state $\left| 1\right\rangle _{E_{1}}$ ($\left|
1\right\rangle _{E_{2}})$. \ Energy must be conserved in the pair-emission
process, and this is indicated by the energy subscript of the one-photon
Fock state for the second photon $\left| 1\right\rangle _{E_{2}=E_{0}-E_{1}}$%
. \ The $integral$ over the product states\ $\left| 1\right\rangle
_{E_{1}}\left| 1\right\rangle _{E_{2}=E_{0}-E_{1}}$indicates that the total
state $\left| \psi \right\rangle $\ is the $superposition$ of product
states. \ Hence, this state exhibits mathematical $nonfactorizability$, the
meaning of which is physical $nonseparability$: It is an \textit{entangled
state of energy}. \ Therefore, the results of measurements of the energy of
the first photon will be tightly correlated with the outcome of measurements
of the energy of the second photon, even when the two photons are
arbitrarily far away from each other. \ There result Einstein-Podolsky-Rosen
effects, which are nonclassical and nonlocal \cite{KwiatThesis},\cite%
{Franson}.

\begin{figure}[tbp]
\centerline{\includegraphics{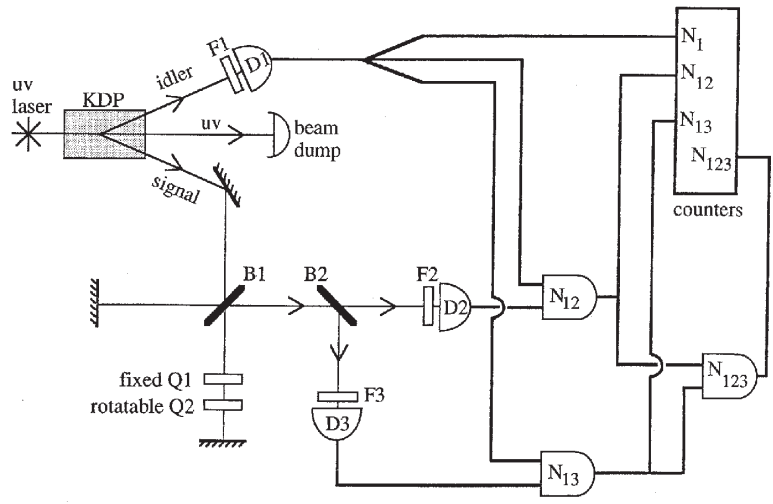}}
\caption{Schematic of apparatus to demonstrate another aspect of
Heisenberg's ``collapse of the wavepacket,'' in which a sharp measurement
of\ the energy of one member of an entangled state results in a collapse in
the width of the wavepacket of the other member. \ A photon pair selected by
means of pinholes shown in Figure 1 (not shown here), is emitted from the
KDP crystal in parametric down-conversion of a parent photon from the UV
laser. \ One member of the pair (the ``idler'') is sent through the
``remote'' interference filter F1 before detection by a photomultiplier D1.
\ The other member of the pair\ (the ``signal'') is sent through a Michelson
interferometer, whose\ optical path length difference is scanned\ by means
of two quarter-wave plates Q1 and Q2. \ The beam splitter B2, filters F2 and
F3, photomultipliers D2 and D3, coincidence gates N$_{12}$ and N$_{13}$, all
serve to select out only those photons which are members of entangled pairs,
for observation. \ The triple coincidence gate N$_{123}$ serves to check
that only single-photon Fock states are detected, so that no classical
explanation of the results would be possible. \ (From Reference \protect\cite%
{Kwiat1991}.)}
\label{fig2}
\end{figure}

\section{Apparatus for the detection of entangled photon pairs: Michelson
interferometry, spectral filtering, and coincidence counting}

In Figure 2, we show a schematic of the apparatus. \ Entangled photons,
labeled ``signal'' and ``idler,'' were produced in the KDP crystal. \ The
upper beam of idler photons was transmitted through the ``remote'' filter F1
to the detector D1, which was a cooled RCA C31034A-02 photomultiplier tube.
Horizontally polarized signal photons in the lower beam entered a Michelson
interferometer, inside one arm of which was placed a pair of zero-order
quarter-wave plates Q1 and Q2. \ The fast axis of Q1 was fixed at 45$^{%
\mathrm{o}}$ to the horizontal, while the fast axis of Q2 was slowly rotated
by a computer-controlled stepping motor, in order to scan for fringes. \
After leaving the Michelson, the signal photon impinged on a second
beamsplitter B2, where it was transmitted to detector D2 through filter F2,
or reflected to detector D3 through filter F3. \ Filters F2 and F3 were
identical: They both had a broad bandwidth of 10 nm centered at $\lambda =$
702 nm. \ Detectors D2 and D3 were identical to D1.\ \ 

Coincidences between detectors D1 and D2 and between D1 and D3 were detected
by feeding their outputs into constant fraction discriminators and
coincidence detectors after appropriate delay lines. \ We used EGG C102B
coincidence detectors with coincidence window resolutions of 1.0 ns and 2.5
ns, respectively. \ Also, triple coincidences between D1, D2, and D3 were
detected by feeding the outputs of the two coincidence counters into a third
coincidence detector (a Tektronix 11302 oscilloscope used in a counter
mode). \ The various count rates were stored on computer every second.

The two quarter-wave plates Q1 and Q2 inside the Michelson generated a
geometrical (Pancharatnam-Berry) phase, which was proportional to the angle
between the fast axes of the two plates. \ Here, one should view the use of
the quarter-wave plates as simply a convenient method for scanning the phase
difference of the Michelson interferometer. \ For the details concerning the
geometrical phase, see \cite{Kwiat1991} and \cite{KwiatThesis}.

We took data both inside and outside the white-light fringe region of the
Michelson (i.e., where the two arms have nearly equal\ optical path
lengths), where the usual interference in singles detection occurs. \ We
report here only on data taken outside this region, where the optical path
length difference was set at a fixed value much greater than the coherence
length (or wavepacket width) of the signal photons determined by the filters
F2 and F3. \ Hence, the fringe visibility seen by detectors D2 and D3 in
singles detection was essentially zero.

\section{Theory}

We present here a simplified analysis of this experiment. \ For a detailed
treatment based on Glauber's correlation functions, see \cite{KwiatThesis}.
\ The entangled state of light after passage of the signal photon through
the Michelson is given by%
\begin{equation}
\left| \psi \right\rangle _{out}=\frac{1}{2}\int dE_{1}A(E_{1})\left|
1\right\rangle _{E_{1}}\left| 1\right\rangle _{E_{2}=E_{0}-E_{1}}\left(
1+\exp (i\phi (E_{0}-E_{1})\right) \text{ ,}
\end{equation}%
where $\phi (E_{0}-E_{1})$ is the total phase shift of the signal photon
inside the Michelson, including the Pancharatnam-Berry phase; the factor of $%
1/2=(1/\sqrt{2})^{2}$ comes from the two interactions with the Michelson
50/50 beam splitter B1. \ The coincidence count rates $N_{12}$ between the
pair of detectors D1 and D2 (and $N_{13}$ between the pair D1 and D3)\ are
proportional to the probabilities of finding at the same instant $t$ (more
precisely, within the detection time, typically 1 ns) one photon at detector
D1 and also one photon at detector D2 (and for $N_{13}$, one photon at
detector D1 and also one photon at detector D3). \ When a $narrowband$
filter F1 centered at energy $E_{1_{0}}$ is placed in front of the detector
D1, $N_{12}$ becomes proportional to%
\begin{equation}
\left| \psi _{out}^{\prime }(\mathbf{r}_{1},\mathbf{r}_{2},t)\right|
^{2}=\left| \left\langle \mathbf{r}_{1},\mathbf{r}_{2},t\right| \left. \psi
\right\rangle _{out}^{\prime }\right| ^{2}\propto 1+\cos \phi \text{ ,}
\end{equation}%
where $\mathbf{r}_{1}$ is the position of detector D1, $\mathbf{r}_{2}$ is
the position of detector D2, and the prime denotes the output state after a
von-Neumann projection onto the eigenstate associated with the sharp energy $%
E_{1_{0}}$ upon measurement. \ Therefore, the phase $\phi $ is determined at
the sharp energy $E_{0}-E_{1_{0}}$. \ In order to conserve total energy, the
energy bandwidth of the collapsed signal photon wavepacket must depend on
the bandwidth of the filter F1 in front of D1, through which it did not
pass. \ Therefore, the visibility of the signal photon fringes seen in
coincidences should depend critically on the bandwidth of this $remote$
filter. \ For a narrowband F1, this fringe visibility should be high,
provided that the optical path length difference of the Michelson does not
exceed the coherence length of the $collapsed$ wavepacket (recall that due
to the energy-time uncertainty principle, collapsing to a $narrower$ energy
spread actually leads to $longer$ wavepacket). \ It should be emphasized
that the width of the collapsed signal photon wavepacket is therefore
determined by the $remote$ filter F1, \textit{through which this signal
photon has apparently never passed!} \ If, however, a sufficiently broadband
remote filter F1 is used instead, such that the optical path length
difference of the Michelson is much greater than the coherence length of the
collapsed wavepacket, then the coincidence fringes should disappear. 
\begin{figure}[tbp]
\centerline{\includegraphics{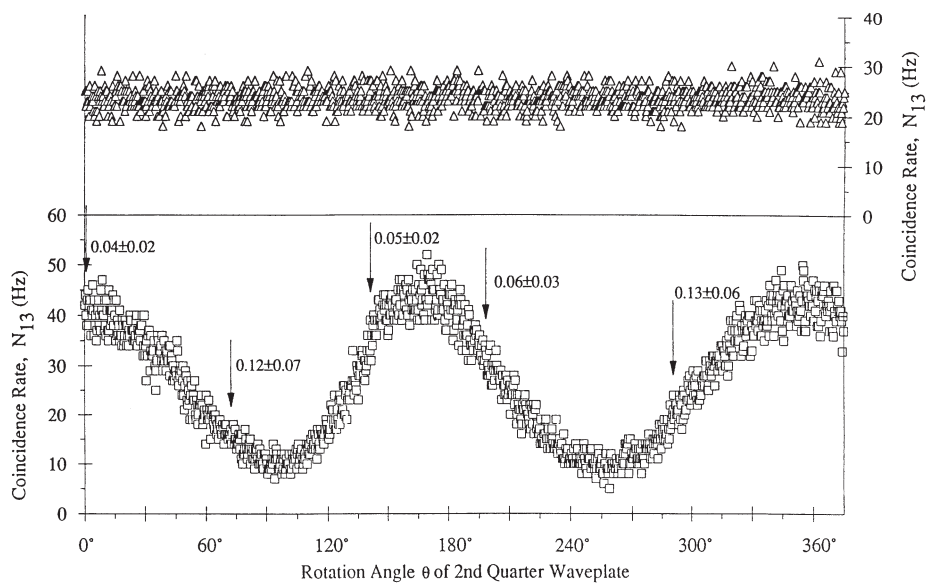}}
\caption{Data demonstrating the phenomenon of the ``collapse of the
wavepacket.'' The visibility of the fringes from the Michelson
interferometer seen in coincidence detection for the signal photon (see
Figure 2), depends on the bandwidth of the $remote$ filter F1, \textit{%
through which it has apparently never passed}. \ In the lower trace, F1 is
narrowband, which results in the collapsed signal photon wavepacket having a
long coherence length, and thus in the observed high-visibility fringes. \
However, when the remote, narrowband filter F1 is replaced by a broadband
one, the collapsed signal photon wavepacket now has a short coherence length
(shorter than the optical path length difference of the Michelson), so that
the fringes disappear, as is evident in the upper trace. \ (Data from
Reference \protect\cite{Kwiat1991}).}
\label{fig3}
\end{figure}

\section{Results}

In Figure 3, we show data which confirm these predictions. \ In the lower
trace (squares) we display the coincidence count rate between detectors D1
and D3, as a function of the angle $\theta $ between the fast axes of the
wave plates Q1 and Q2, when the remote filter F1 was quite narrowband (i.e.,
with a bandwidth of 0.86 nm). \ The calculated coherence length of the
collapsed signal photon wavepacket (570 microns) was greater than the
optical path length difference at which the Michelson was set (220 microns).
\ The observed visibility of the coincidence fringes was quite high, viz.,
60\% $\pm $ 5\%, indicating that the collapse of the signal photon
wavepacket had indeed occurred.\

In the upper trace (triangles) we display the coincidence count rate versus $%
\theta $ when a broadband remote filter F1 (i.e., one with a bandwidth of 10
nm) was substituted for the narrowband one. \ The coherence length of the
collapsed signal photon wavepacket in this case should have been only 50
microns, which is shorter than the 220 micron optical path length difference
at which the Michelson was set. \ The observed coincidence fringes have now
indeed disappeared, indicating that the collapse of the signal photon
wavepacket (this time to a $shorter$ temporal width) has again occurred.

In addition to the above coincidences-counting data, we also took
singles-counting data at detector D3 (where are not shown here), at the same
settings of the optical path length of the Michelson for the traces shown in
Figure 3. \ We observed no visible fringes in the singles-counting data
during the scan of $\theta $, with a visibility less than 2\%. \ This
indicates that we were well outside of the white-light fringe of the
Michelson. \ More importantly, this also demonstrates that only those
photons which are detected in $coincidence$ with their twins which have
passed through the narrowband filter F1, exhibit the observed phenomenon of
the collapse of the wavepacket upon detection. \ In other words, only $%
entangled$ pairs of photons detected by\ the coincidences-counting method
show this kind of ``collapse'' behavior.

Heisenberg's energy-time uncertainty principle was also demonstrated during
the course of this experiment \cite{NASA}. \ The width $\Delta t_{2}$ of the
collapsed signal photon wavepacket, which was measured by means of the
Michelson, satisfied the inequality%
\begin{equation}
\Delta E_{2}\Delta t_{2}\geq \hbar /2\text{ ,}  \label{uncertainty}
\end{equation}%
where the energy width $\Delta E_{2}$ of the collapsed signal photon
wavepacket, was determined by the measured energy width $\Delta E_{1}$ of
the idler photon, in order to conserve total energy. \ Hence, the energy
width $\Delta E_{2}$ of the signal photon, which enters into the Heisenberg
uncertainty relation (\ref{uncertainty}), was actually the width $\Delta
E_{1}$ of the remote filter F1, through which this signal photon did not
pass.

\section{Discussion}

Any classical electromagnetic field explanation of these results can be
ruled out. \ We followed here the earlier experiment of Grangier, Roger, and
Aspect \cite{Grangier}, in which they showed that one can rule out any
classical-wave explanation of the action of an optical beam splitter, by
means of triple coincidence measurements in a setup similar to that shown in
Figure 2. \ Let us define the parameter 
\begin{equation}
a=\frac{N_{123}N_{1}}{N_{12}N_{13}}\text{ ,}
\end{equation}%
where $N_{123}$ is the count rate of triple coincidences between detectors
D1, D2, and D3, $N_{12}$ is the count rate of double coincidences between
detectors D1 and D2, $N_{13}$ is the count rate of double coincidences
between detectors D1 and D3, and $N_{1}$ is the count rate of singles
detection by D1 alone. \ From Schwartz's inequality, it can be shown that
for any classical-wave theory for electromagnetic wavepackets, the
inequality $a\geq 1$ must hold \cite{Grangier}. \ By contrast, quantum
theory predicts that $a=0$. \  

The physical meaning of the inequality $a\geq 1$ is that any classical
wavepacket divides itself smoothly at a beam splitter, and this always
results in triple coincidences after the beam splitter B2, in conjunction
with a semi-classical theory of the photoelectric effect. \ By contrast, a
single photon is indivisible, so that it cannot divide itself at the
beamsplitter B2, but rather must exit through only one or the other of the
two exit ports of the beam splitter, and this results in zero triple
coincidences (except for a small background arising from multiple pair
events).

For coincidence-detection efficiencies $\eta $ less than unity, this
inequality becomes $a\geq \eta $. \ We calibrated our triple-coincidence
counting system by replacing the photon-pair light source by an attenuated
light bulb, and measured $\eta =70\%\pm 7\%$. \ During the data run of
Figure 3 (lower trace), we measured values of $a$ shown at the vertical
arrows. \ The average value of $a$ is $0.08\pm 0.04$, which violates by more
than thirteen standard deviations the predictions based on any
classical-wave theory of electromagnetism.

It is therefore incorrect to explain these results in terms of a
stochastic-ensemble model of classical electromagnetic waves, along with a
semi-classical theory of photoelectric detection \cite{Clauser}. \ Pairs of
classical waves with conjugately correlated, but random, frequencies could
conceivably yield the observed interference patterns, but they would also
yield many more triple coincidences than were observed.

One might be tempted to explain our results simply in terms of conditioning
on the detection of the idler photons to post-select signal photons of 
\textit{pre-existing definite} energies. And in fact, such a local realistic
model can account for the results of this experiment with no need to invoke
a nonlocal collapse. However, in light of the observed violations of Bell's
inequalities based on energy-time variables \cite{Franson}, it is incorrect
to interpret these results in terms of a statistical ensemble of signal and
idler photons which possess definite, but unknown, energies before
measurement (i.e., before filtering and detection). \ Physical observables,
such as energy and momentum, cannot be viewed as local, realistic properties
which are carried by the photon during its flight to the detectors, before
they are actually measured.

We have chosen to interpret our experiment in terms of the ``collapse''
idea. \ However, it should be stressed that this is but one interpretation
of quantum measurement. \ Other interpretations exist which could possibly
also explain our results. \ They include the Bohm-trajectory picture \cite%
{Bohm}, the many-worlds interpretation \cite{Everitt}, the advanced-wave or
transactional model by Cramer \cite{Cramer}\ (and related ideas by Klyshko %
\cite{Klyshko}), the ``non-collapse'' quantum-cosmology picture \cite{Hartle}%
, and others.

An obvious follow-up experiment (but one which has not yet been performed)
is a version of Wheeler's ``delayed-choice'' experiment \cite{Wheeler}, in
which one could increase, as much as one desires, the distance from the
source to the filter F1 and the detector D1 on the ``remote'' side of the
apparatus, as compared to the distance to detectors D2 and D3, etc., on the
``near'' side. \ The arbitrary choice of whether filter F1 should be
broadband or narrowband could then be delayed by the experimenter until well
after ``clicks'' had already been irreversibly registered in detectors D2 or
D3. \ In this way, we can be sure that the signal wavepacket on the near
side of the apparatus could not have known, well in advance\ of the
experimenter's arbitrary and delayed choice of F1 on the remote side of the
apparatus, whether to have collapsed to a broad or to a narrow wavepacket. \ 

However, there is no paradox here since the determination of coincident
events can only be made $after$ the records of the ``clicks'' at both near
and remote detectors are brought together and compared, and only then does
the appearance or disappearance of interference fringes in coincidences
become apparent. \ The bringing together of these records requires the
propagation of signals through \textit{classical channels} with appropriate
delays, such as the post-detection coaxial delay lines leading to the
coincidence gates shown in Figure 2. \ Classical channels propagate signals
with discontinuous fronts, such as ``clicks,'' at a speed limited by $c$. \
Hence there is no conflict with the postulates of the theory of relativity.
\  

It is\ therefore incorrect to say that the experimenter's arbitrary choice
of the filter F1 on the remote side of the apparatus somehow $caused$ the
collapse of the signal photon wavepacket on the near side. \ Only nonlocal,
instantaneous, \textit{uncaused correlations-at-a-distance} are predicted by
quantum theory. \ Clearly, the collapse phenomenon is \textit{nonlocal and
noncausal} in nature.

In conclusion, we have demonstrated that the $nonlocal$ collapse of the
wavefunction or wavepacket in the Copenhagen interpretation of quantum
theory, which was introduced by Heisenberg in 1929, leads to a
self-consistent description of our experimental results. \ Whether or not
fringes in coincidence detection show up in a Michelson interferometer on
the near side of the apparatus, depends on the arbitrary choice by the
experimenter of the remote filter F1 through which the photon on the near
side has evidently never passed. \ This collapse phenomenon, however, is
clearly $noncausal$, as a ``delayed-choice'' extension of our experiment
would show.

\section{Acknowledgments}

This work was supported in part by the ONR. \ We thank Charles H. Townes for
a careful reading of this manuscript. \ RYC would like to thank John C.
Garrison for helpful discussions, and Aephraim M. Steinberg for pointing out
to me once again Reference [1].


\begin{thebibliography}{99}
\bibitem{Heisenberg1929} \textit{The Physical Principles of the Quantum
Theory}, W. Heisenberg (University of Chicago Press, Chicago, 1930),
reprinted by Dover Publications, p. 39.

\bibitem{Born} M. Born, Z. f. Physik \textbf{37}, 863 (1926); see also the
commentary by L. Rosenfeld in \textit{Quantum Theory and Measurement},
edited by J. A. Wheeler and W. H. Zurek (Princeton University Press,
Princeton, 1983), p. 50.

\bibitem{HeisenbergFootnote} Heisenberg added the following footnote here:
``For a single photon the configuration space has only three dimensions; the
Schr\"{o}dinger equation of a photon can thus be regarded as formally
identical with the Maxwell equations.''

\bibitem{Newton} It is interesting to note here that much earlier, Newton,
in connection with his experimental studies of the optical beam splitter (or
what Heisenberg called a ``semi-transparent mirror''), and in particular, of
the phenomenon of Newton's rings, struggled with the problem of how to
reconcile his concept of an indivisible ``corpuscle'' (i.e., a particle) of
light, with the\ fact that these corpuscles were $coherently$ divided into
transmitted and reflected parts at air-glass interfaces, in such a way that
these parts could later interfere with each other, and\ thus produce
Newton's rings.

\bibitem{Dicke} R. H. Dicke, Am. J. Phys. \textbf{49}, 925 (1981); A. C.
Elitzur and L. Vaidman, Found. Phys. \textbf{23}, 987 (1993); P. G. Kwiat,
H. Weinfurter, T. Herzog, A. Zeilinger, and M. A. Kasevich, Phys. Rev. Lett. 
\textbf{74}, 4763 (1995).

\bibitem{Von Neumann} \textit{Mathemathical Foundations of Quantum Mechanics}%
, J. von Neumann (Princeton University Press, Princeton, 1955), p. 347ff.

\bibitem{Burnham} S. E. Harris, M. K. Oshman, and R. L. Byer, Phys. Rev.
Lett. \textbf{18}, 732 (1967); D. Magde and H. Mahr, Phys. Rev. Lett. 
\textbf{18}, 905 (1967); D. C. Burnham and D. L. Weinberg, Phys. Rev. Lett. 
\textbf{25}, 84 (1970).

\bibitem{Kwiat1991} P. G. Kwiat and R. Y. Chiao, Phys. Rev. Lett. \textbf{66}%
, 588 (1991).

\bibitem{CWpump} Strictly speaking, because we have a $continuous$ (CW)
pump, before detecting one of the down-conversion photons we cannot describe
them as wavepackets. It is only $after$ detecting one photon that its sister
photon is in a proper wavepacket.

\bibitem{OmitVacuum} We omit here the predominant contribution of the
vacuum, corresponding to events where the pump photon did $not$ down
convert; such events are never measured in a photon counting experiment such
as ours. We also omit the entirely negligible contribution from
multiple-pair events (i.e., in which more than one pair was produced within
the detection time).

\bibitem{KwiatThesis} P. G. Kwiat, Ph. D. Thesis, U. C. Berkeley (1993,
unpublished but available at \TEXTsymbol{<}%
http://wwwlib.umi.com/dxweb/details?doc\_no=2717843\TEXTsymbol{>}).

\bibitem{Franson} J. D. Franson, Phys. Rev. Lett. \textbf{62}, 2205 (1989);
J. Brendel, E. Mohler, and W. Martienssen, Europhys. Lett. \textbf{20}, 575
(1992); P. G. Kwiat, A. M. Steinberg, and R. Y. Chiao, Phys. Rev. A \textbf{%
47}, R2472 (1993); S. Aerts, P. Kwiat, J.-\AA . Larsson, and M. $\mathrm{{%
\dot{Z}}}$ukowski, Phys. Rev. Lett. \textbf{83}, 2872 (1999).

\bibitem{NASA} R. Y. Chiao, P. G. Kwiat, and A. M. Steinberg, in the \textit{%
Workshop on Squeezed States and Uncertainty Relations}, D. Han et al., eds.,
NASA Conference Publication 3135, NASA, Washington, DC, 1991, p. 61.

\bibitem{Grangier} P. Grangier, G. Roger, and A. Aspect, Europhys. Lett. 
\textbf{1}, 173 (1986).

\bibitem{Clauser} J. F. Clauser, Phys. Rev. D \textbf{9}, 853 (1974).

\bibitem{Bohm} D. Bohm, Phys. Rev. \textbf{85}, 166 (1951); \textbf{85}, 180
(1952); D. Bohm, B. J. Hiley, and P. N. Kaloyerou, Phys. Rep. \textbf{144},
321 (1987).

\bibitem{Everitt} H. Everitt, Rev. Mod. Phys. 29, 454 (1957); H. Everitt, 
\textit{The Many-Worlds Interpretation of Quantum Mechanics} (Princeton
University Press, Princeton, NJ, 1973); L. Vaidman `The Many-Worlds
Interpretation of Quantum Mechanics,' to appear in \textit{The Stanford
Encyclopedia of Philosophy}, available at \TEXTsymbol{<}http://www.tau.ac.il/%
\symbol{126}vaidman/mwi/mw2.html\TEXTsymbol{>}.

\bibitem{Cramer} J. G. Cramer, Rev. Mod. Phys. \textbf{58}, 647 (1986); Int.
J. Th. Phys. \textbf{27}, 227 (1988); `The Plane of the Present and the New
Transactional Paradigm of Time' in \textit{Time and the Instant}, Robin
Durie, ed., (Clinamen Press, Manchester, 2000), Chap. 8.

\bibitem{Klyshko} D. N. Klysko, Phys. Lett. A \textbf{132}, 299 (1988).

\bibitem{Hartle} M. Gell-Mann and J. B. Hartle `Quantum mechanics in light
of quantum cosmology,' in \textit{Complexity, Entropy and the Physics of
Information}, W. H. Zurek, ed. (Addison-Wesley, Reading, 1990), p. 425;
Phys. Rev. D \textbf{47}, 3345 (1993). 

\bibitem{Wheeler} J. A. Wheeler in \textit{Quantum Theory and Measurement},
edited by J. A. Wheeler and W. H. Zurek (Princeton University Press,
Princeton, 1983), p. 182.
\end{thebibliography}
\end{document}